\documentclass[a4paper]{article}
\usepackage{epsfig}
\begin{document}
\large

{\parindent=0pt
{\Large  \bf
\begin{center}
Vibration of the Dimer on Si(001) Surface \\
Excited by STM Current
\end{center}
}
}


{\parindent=0pt
\begin{center}
Hiroshi Kawai$^*$, Osamu Narikiyo
\end{center}
}


{\parindent=0pt
{\normalsize 
\begin{center}
 Department of Physics, Faculty of Sciences, Kyushu University, \\
Ropponmatsu, Fukuoka 810-8560, Japan\\
\end{center}
}
}

\begin{center}
(Received \hspace*{8cm})
\end{center}

\vspace*{1cm}

{\parindent=0pt \bf  Abstract}

The vibration of the dimer excited by STM current on Si(001) surface
is investigated. 
We describe this system by 
the Hamiltonian which has the electron-vibration coupling term 
as the key ingredient.
In order to characterize the transition rates induced by STM current
between vibrational states
we have introduced the effective temperature of the vibration
which differs from the temperature of the substrate.
The behavior of the effective temperature depends on
the substrate temperature and STM current in highly nonlinear manner and
qualitatively changes around 50K of the substrate temperature.
At lower temperatures, the effective temperature strongly deviates
from the substrate temperature and
reaches a few hundred Kelvin for the typical values of STM current.
At higher substrate temperatures, the effective temperature 
reduces to the substrate temperature.
On the basis of these behaviors of the effective temperature,
we solve the puzzle of the symmetric-asymmetric crossover in dimer
images of STM observation in the ordered state of c(4 $\times$ 2).
\\

\vspace*{1cm}

{\parindent=0pt 
{\bf KEYWORDS}: Si(001), symmetric image, STM, dimer vibration

\vspace*{4cm}

$^*$ E-mail: kawai@rc.kyushu-u.ac.jp
}

\newpage

{\parindent=0pt \bf 1. Introduction}

Intensive studies have been made on the atomic structure of the reconstructed 
Si(001) surface both experimentally and theoretically.
On Si(001) surface, neighboring atoms 
form buckled dimers which have two stable 
tilting angles. Along the 
$\langle 110 \rangle$ direction, the dimer
rows is formed by the dimers.
The principal features of the surface reconstruction
on Ge(001) surface are similar to Si(001) surface.
The ground state of the structure of the dimer arrangement of both surfaces is 
c(4 $\times$ 2) and 
the ordered c(4 $\times$ 2) phase turns into a disordered state
at room temperature
through the order-disorder phase transition.$^{1-14)}$ 
In the ordered state, the dimers are observed as the asymmetric images
by scanning tunneling
microscopy (STM).
In the disordered state, the symmetric appearing images are observed by STM,
and these symmetric appearing images have been attributed
to the rapid repeat of the orientational change (flip-flop motion).

In the STM observation, semiconductor substrates are doped into n-type ones
or p-type ones for the electric conductivity. 
On Si(001) surface,
serious differences 
in the property of the dimer system
between the n-type substrates and 
p-type ones have not been  reported 
at temperatures higher than 60K.

Recently, p(2 $\times$ 2) phase on Si(001) surface$^{15-17)}$ 
is observed on 
highly doped n-type substrates below 40K by STM. 
A phase transition is also observed by
low-energy electron diffraction$^{18)}$ 
(LEED) at 40K on 
highly doped n-type substrates. 
On the p-type substrates, 
c(4 $\times$ 2) phase  on Si(001) surface is observed down to 9K.$^{17)}$ 

At much low temperatures, the symmetric STM images of p(2 $\times$ 1)
structure$^{17,19,20)}$ 
are observed
at some conditions of the tip bias voltages $V_{\rm t}$ and the tip
currents $I_{\rm t}$ in STM observations.
Yokoyama$^{19)}$ 
reported that
STM images of the dimers at 5K on boron doped (B-doped) 
p-type Si(001) substrate
at the condition of $V_{\rm t}= \pm 1$V and $I_{\rm t}=50$pA   
appear to be symmetric, 
although most of the dimer images on the substrate are well asymmetric
to make the ordered c(4 $\times$ 2) structure at 63K,
and that the symmetric STM images at 5K are 
observed also on Antimony doped n-type Si(001) substrate.
Kondo$^{19)}$ 
 reported subsequently that
the STM images at 20K on  phosphorus doped n-type Si(001) substrate
at the condition of $V_{\rm t}= 2.0$V and -1.6V, and $I_{\rm t}=300$pA   
appear to be symmetric.
These observations of the symmetric images in the ordered phase
are puzzling and have not been explained yet.

The first-principles calculations (FPC) with
high accuracy$^{6,7,21)}$ 
show that 
the ground state  on Si(001) surface is c(4 $\times$ 2).
The results of FPC show that
the energy of the electronic system
decreases, and
the energy of the lattice system increases, 
as the tilting angle of the dimer increases.  
The equilibrium angle is determined
by the competition between the energies of two systems. 
Thermal equilibrium
properties on the surface at temperatures higher than 100K have 
been successfully explained by 
Monte Carlo simulations (MCS) with  Metropolis algorithm based
on the results of FPC$^{6,7)}$ 
by mapping the two stable orientations of the dimer
onto the Ising system.$^{6,7,22-25)}$ 
The time-resolving dynamical Monte Carlo simulations (TDMCS)
performed for the dimer systems of Si(001)$^{26,27)}$ 
surface and Ge(001)$^{28)}$ 
surface
show that
the most of the long range fluctuations
of the dimer arrangement on Si(001) and Ge(001) surfaces
are carried by the phase boundary of the c(4 $\times$ 2)
domain in the dimer row  at low temperatures.
In order to perform TDMCS on Si(001) surface and 
on Ge(001) surface, the model potentials for the continuous
values of the tilting angle are obtained based on the 
FPC results$^{27,28)}$. 
The TDMCS results with the model potentials
reproduce well the results of the time-resolving STM 
observations$^{12,29)}$. 
Since at temperatures higher than 60K, as mentioned above, 
serious dopant effect are not reported on Si(001) surface,
the reported MCS results based on the results of FPC remain valid.

The surface localized states of $\pi$-band 
and $\pi^*$-band$^{6,7,17,19,21,30-32)}$ 
are formed mainly through the hybridization of 
the dangling bonds on the dimers.
FPC$^{6,7,21)}$ 
show that the energies of the top of the $\pi$-band and 
the bottom of $\pi^*$-band stay a few ten meV above from the top of 
the valence band of Si and about 500meV below from the bottom
of the conduction band of Si, respectively.   
The band widths are obtained to be about 1eV in both of
$\pi$-band and $\pi^*$-band. 
These features are observed by angle-resolved UPS$^{30)}$   
(ARUPS) and   
scanning tunneling spectroscopy $^{19,31,32)}$ 
(STS).
On the n-type substrates, the Fermi level of the system stays
near the bottom of the conduction band,
which is much higher than the bottom of $\pi^*$-band.
The large band bending appears near the surface region
of the n-type substrate at much low temperatures.
On the p-type substrates, the Fermi level stays    
near the top of the valence band.
Therefore, the top of the $\pi$-band is near the Fermi level.
The large band bending does not appear near the surface region 
on the p-type substrates.
Because 
the dopant effect on the substrate is not taken into account
in the FPC, the results of FPC 
reproduce well the surface properties of p-type substrate,
but some modifications are necessary for n-type substrate
at much low temperatures.

In the present study, we take the scope to 
the vibration excited by STM current on 
the p-type substrate.  
We will introduce the Hamiltonian 
which has the electron-vibration coupling term.
The excitation rates by the STM currents for the vibration
will be obtained.
The rates will be characterized by introducing
the effective temperature of the vibration.
The mechanism for the symmetric-asymmetric crossover in
dimer images
of STM observation on Si(001) surface will be presented.\\
 
{\parindent=0pt \bf 2. Model}

We introduce the Hamiltonian  $H$ 
representing the coupling between the electronic system and 
the localized vibrational system,

\begin{eqnarray}
        H &=& H_{\rm e} + H_{\rm e-v}, \nonumber \\
H_{\rm e} &=& \sum_k \varepsilon_k c_k^\dagger c_k 
            +  \sum_p \varepsilon_p c_p^\dagger c_p
   + \Gamma \left\{(\sum_k \gamma_k c_k^\dagger)(\sum_p \gamma_p^* c_p) + {\rm H.c.} 
\right\} \nonumber \\
          &=& \sum_\alpha \varepsilon_\alpha c_\alpha^\dagger c_\alpha
            + \varepsilon_a a^\dagger a
            + (\sum_\alpha \Gamma_\alpha a^\dagger c_\alpha + {\rm H.c.})
            + \sum_p \varepsilon_p c_p^\dagger c_p
            + (\sum_p \Gamma_p a^\dagger c_p + {\rm H.c.}) \nonumber, \\
H_{\rm e-v}&=& \hbar \omega (b^\dagger b + \frac{1}{2}) 
            + \delta \varepsilon (b^\dagger +b) a^\dagger a, \nonumber \\
\varepsilon_a &=& \sum_k |\gamma_k|^2 \varepsilon_k, \nonumber \\
    a^\dagger &=& \sum_k \gamma_k c_k^\dagger, \nonumber \\
    \Gamma_p  &=& \Gamma \gamma_p,
\end{eqnarray} 
where $c_k$, $c_p$ and $b$ are the annihilation operators for the
electronic states 
in the $\pi$-band,
for the states in the conduction band of the STM tip, and for the vibrational
state in the rocking mode of the dimer on Si(001) surface beneath
the STM tip. We assume that the vibrational
states are the harmonic oscillator, and 
the energy of the state with the vibrational number $n$ is 
$\hbar \omega (n+1/2)$.
The energy of the vibrational state $\hbar \omega$ is measure by 
the electron energy loss spectroscopy$^{33)}$ 
as 20meV.
$a^\dagger = \sum_k \gamma_k c_k^\dagger$ and 
$\sum_p \gamma_p c_p^\dagger$ are the creation operator for
the spatially localized states $|a \rangle$
in the surface dimer
and in the prominence of the tip, respectively.
The tunneling current is through the the spatial localized states.
The coefficients of the spatial localized states are normalized 
as $\sum_k |\gamma_k|^2=1$ and  $\sum_p |\gamma_p|^2=1$.
The diagonalized term of $\sum_k \varepsilon_k c_k^\dagger c_k$
is rewritten by $a$ and $c_\alpha$ 
as the first three terms in the third line of eq. (1), 
where $c_\alpha$ is the annihilation operators for the
electronic states orthogonalized with $|a \rangle$.
The amplitude of the vibration  modifies the one electron
energy of $|a \rangle$, and  $H$ has the electron-vibration coupling term,
because the energy of the electronic system, as mentioned above,
decreases as the tilting angle of the dimer increases.
The term of 
$\delta \varepsilon$ represents
the electron-vibration coupling.
$\delta \varepsilon$ is the coupling constant between
the amplitude of the vibration and the energy of the one electron state
of $|a \rangle$. 
The vibration is excited or deexcited through 
the electron-vibration coupling by the STM current. 
The similar Hamiltonian of eq. (1) has been used in
the precedent studies
of vibrations of molecules adsorbed on metal 
surfaces$^{34-41)}$. 
In the present study, however, the Hamiltonian is applied to 
totally different system, the surface
localized semiconductor state on Si(001) surface.

In the present study, 
we treat the case that the tip bias voltages $V_{\rm t}$ is
positive and so large that the Fermi level in the conduction band
of the STM tip is below the bottom of the $\pi$-band of 
the p-type Si(001) substrate;
$V_{\rm t}$ is assumed to be slightly larger than 1V, 
because the band width of the $\pi$-band is about 1eV. 
The elastic inter-band transition of electrons from the  $\pi$-band 
to the conduction band of the tip 
is induced
by the presence of the tip. 
The elastic inter-band transition rate $\sigma_{\rm elas}$ of electrons from the surface
to the tip without coupling to the vibration$^{34-41)}$ is given
by the T-matrix element and the Fermi distribution functions:

\begin{eqnarray}
\sigma_{\rm elas} &=& 2\frac{2 \pi}{\hbar}\sum_{\alpha, p}
|\Gamma_p^* G_a(\varepsilon) \Gamma_\alpha|^2  \nonumber \\
&& \quad {} \times 
(\frac{1}{\exp(\beta(\varepsilon_\alpha-\varepsilon_{\rm F}))+1})
(1-\frac{1}{\exp(\beta(\varepsilon_p+eV_{\rm t}-\varepsilon_{\rm F}))+1})
\delta(\varepsilon_\alpha-\varepsilon_p) \nonumber \\
&=& \frac{4}{\pi \hbar} 
\int_{-D_\pi+E_\pi}^{E_\pi} {\rm d} \varepsilon \> 
\Delta_{\rm t} (\varepsilon)
\Delta_{\rm s} (\varepsilon)
|G_a(\varepsilon)|^2  \nonumber \\
&& \quad {} \times 
(\frac{1}{\exp(\beta(\varepsilon_\alpha-\varepsilon_{\rm F}))+1})
(\frac{1}{1+\exp(-\beta(\varepsilon_p+eV_{\rm t}-\varepsilon_{\rm F}))})
\nonumber \\
&\approx& \frac{4}{\pi \hbar} \Delta_{\rm t} (\varepsilon_a) \Delta_{\rm s} (\varepsilon_a) 
|G_a (\varepsilon_a)|^2 D_\pi, \nonumber \\
G_a (\varepsilon)&=& \langle a | (\varepsilon + i 0^+ - H_{\rm e})^{-1} | a \rangle   
\nonumber \\
&=& \frac{1}{\varepsilon -\varepsilon_a - \Lambda(\varepsilon) + i \Delta (\varepsilon)}, 
\nonumber \\
\Delta (\varepsilon) &=& \Delta_{\rm s} (\varepsilon) +  \Delta_{\rm t} (\varepsilon), 
\nonumber \\
\Delta_{\rm s} (\varepsilon) &=& \pi \sum_\alpha |\Gamma_\alpha|^2 
\delta(\varepsilon-\varepsilon_\alpha),
\nonumber \\
\Delta_{\rm t} (\varepsilon) &=& \pi \sum_p |\Gamma_p|^2 
\delta(\varepsilon-\varepsilon_p),
\nonumber \\
\Lambda (\varepsilon) &=& \frac{1}{\pi} P 
\int_{-\infty}^\infty {\rm d} \varepsilon' 
\frac{\Delta (\varepsilon')}{\varepsilon-\varepsilon'},
\end{eqnarray}  
where $\varepsilon_{\rm F}$ is the Fermi level of the electronic
system, $G_a (\varepsilon)$ is the Green function
for the electron system, $\Delta (\varepsilon)$ is 
the width of the projected density of states for
$|a \rangle$,
$\Delta_{\rm s} (\varepsilon)$ and $\Delta_{\rm t} (\varepsilon)$ 
are the components of $\Delta (\varepsilon)$ in the surface and in the tip, respectively,
$P$ denotes the Cauchy principal value,
and $E_\pi$ and $D_\pi$ are the energy level of the top of 
the $\pi$-band and the width of the $\pi$-band, respectively.
In the present study, $-(E_\pi - \varepsilon_{\rm F})$ is
assumed to be of the order of a few ten meV 
and $D_\pi$ is assumed to be large enough $\beta D_\pi \gg 1$. 
The temperature of the system is assumed to be so low
that the broadness of the Fermi distribution around 
$\varepsilon_{\rm F}$ in the surface state
and around 
$\varepsilon_{\rm F}-eV_{\rm t}$ in the tip state
are not contributed to the evaluation of the integration in eq.(2).
We assume that $\Delta_{\rm t} (\varepsilon_a) / \Delta_{\rm s} (\varepsilon_a)$
is small in the STM observation in the typical
conditions enough to be  approximated  as
$\Delta (\varepsilon_a) \approx \Delta_{\rm s} (\varepsilon_a)$. 
Because in the STS experiment at 5.5K, the tunneling spectra of the $\pi$-band 
are observed to be almost symmetric band, 
we assume that $\varepsilon_a$ stay at the center of the
$\pi$-band in energy, and $\Lambda (\varepsilon_a)=0$. 
Therefore, $\sigma_{\rm elas}$ is approximately obtained as 
     
\begin{eqnarray} 
\sigma_{\rm elas} &\approx& 
\frac{4}{\pi \hbar}
\Delta_{\rm t} (\varepsilon_a) \Delta_{\rm s}^{-1} (\varepsilon_a) D_\pi .
\end{eqnarray}   

The inter-band transition of electrons from the $\pi$-band 
to the conduction band of the tip 
induced
by the presence of the tip
can excite the vibrational state in the dimer
through the electron-vibration coupling.
The inelastic inter-band transition  rate $\sigma_{0 \to 1}^{\rm int}$
of electrons$^{34-41)}$ 
from the surface to the tip coupling 
with the excitation from the ground state of the vibrational
number $n=0$ to the excited state of $n=1$ is obtained
in the lowest order in $|\delta \varepsilon|$ as

\begin{eqnarray} 
\sigma_{0 \to 1}^{\rm int} &=& 2 \frac{2\pi}{\hbar} \sum_{\alpha', p'} 
| \langle p', 1 |H_{\rm e-v}|\alpha', 0 \rangle |^2 
\delta(E(p') - \hbar \omega - E(\alpha')) \nonumber \\
        &=& \frac{4\pi |\delta \varepsilon|^2}{\hbar}
          \int_{-D\pi+E_\pi}^{E_\pi} {\rm d} \varepsilon \> 
\rho_{\rm s}(\varepsilon) \rho_{\rm t}(\varepsilon)  \nonumber \\
&& \quad {} \times 
(\frac{1}{\exp(\beta(\varepsilon_\alpha-\varepsilon_{\rm F}))+1})
(1-\frac{1}{\exp(\beta(\varepsilon_p - \hbar \omega +eV_{\rm t}-\varepsilon_{\rm F}))+1}) \nonumber \\
&\approx& \frac{4\pi |\delta \varepsilon|^2}{\hbar} 
\rho_{\rm s}(\varepsilon_a) \rho_{\rm t}(\varepsilon_a)D_\pi, \nonumber \\
\rho_{\rm s}(\varepsilon) &=& \sum_{\alpha'} |\langle \alpha' | a \rangle|^2 
\delta(\varepsilon - \varepsilon_{\alpha'}) \nonumber \\
  &=& \frac{1}{\pi}
(\frac{\Delta_{\rm s}(\varepsilon)}{(\varepsilon-\varepsilon_a-\Lambda(\varepsilon))^2
+\Delta^2(\varepsilon))}), \nonumber \\
\rho_{\rm t}(\varepsilon) &=& \sum_{p'} |\langle p' | a \rangle|^2 
\delta(\varepsilon - \varepsilon_{p'}) \nonumber \\
  &=& \frac{1}{\pi}
(\frac{\Delta_{\rm t}(\varepsilon)}{(\varepsilon-\varepsilon_a-\Lambda(\varepsilon))^2
+\Delta^2(\varepsilon))}), 
\end{eqnarray} 
where $E(\alpha')$ and $E(p')$ are the one electron energies of
$| \alpha' \rangle$ and $| p' \rangle$, respectively, and  
$\rho_{\rm s}$ and $\rho_{\rm t}$ are 
the projected density
of states for $| a \rangle$. 
$| \alpha' \rangle$ and $| p' \rangle$ are 
the stationary states of $H_{\rm e}$ connected to 
$| \alpha \rangle$ and $| p \rangle$, respectively. 
Namely, the stationary states are given as
$|\alpha' \rangle = | \alpha \rangle 
+ 
(\varepsilon_\alpha +i 0^+ - H_{\rm e})^{-1} \Gamma_\alpha | a \rangle $
and
$| p' \rangle = | p \rangle 
+ (\varepsilon_p +i 0^+ - H_{\rm e})^{-1} \Gamma_p | a \rangle$.  
The  projected density
of states are approximated in the same way as eq. (3):
$\rho_{\rm s}(\varepsilon_a) \approx 
(\pi \Delta_{\rm s}(\varepsilon_a))^{-1}$ 
and 
$\rho_{\rm t}(\varepsilon_a) \approx \Delta_{\rm t}(\varepsilon_a)
(\pi \Delta_{\rm s}^2 (\varepsilon_a))^{-1}$. 
The inelastic transition rate $\sigma_{0 \to 1}^{\rm int}$
is given 
by use of the approximation for the projected density of states as

\begin{eqnarray} 
\sigma_{0 \to 1}^{\rm int} &\approx& 
\frac{4 |\delta \varepsilon|^2}{\pi \hbar}
\Delta_{\rm t} (\varepsilon_a) \Delta_{\rm s}^{-3} (\varepsilon_a) D_\pi .
\end{eqnarray}   

The elastic transition rate $\sigma_{\rm elas}$ and 
the inelastic transition rate $\sigma_{0 \to 1}^{\rm int}$
depend  critically on $I_{\rm t}$
and do not on $T$,
because in eqs. (3) and (5), $\Delta_{\rm t}$
depends  critically on the spatial 
tunneling gap of STM.
The fraction of the inelastic transition rate 
to the elastic transition rate$^{34-41)}$  
$W=\sigma_{0 \to 1}^{\rm int}/\sigma_{\rm elas}$, however, does not
depend on $\Delta_{\rm t}$, and is obtained from eqs. (3) and (5) as
 
\begin{eqnarray} 
W &=& 
(\frac{|\delta \varepsilon |}{\Delta_{\rm s}(\varepsilon_a)})^2 .
\end{eqnarray} 
Eq. (6) shows that the fraction  of 
$\sigma_{0 \to 1}^{\rm int}/\sigma_{\rm elas}$
is given dominantly from the intrinsic property of Si(001) surface,
and independent of the STM current.
In the present study, as mentioned already, $V_{\rm t}$ is
assumed
to be so large that the Fermi level in the conduction band
of the tip is below the bottom of the $\pi$-band, and
$\hbar \omega \ll eV_{\rm t}$.
From these assumption, 
the inelastic inter-band transition rate $\sigma_{1 \to 0}^{\rm int}$
of electrons 
from the surface to the tip  coupling 
with the deexcitation from the vibrational
number $n=1$ to $n=0$ is well approximated to take
the same value as
$\sigma_{0 \to 1}^{\rm int}$. 
Thus 
the inelastic inter-band transition  rate $\sigma_{n \to n \pm 1}^{\rm int}$
of electron 
from the surface to the tip coupling 
with the vibrational transition from $n$ to 
$n \pm 1$ is given as 
$\sigma_{n \to n \pm 1}^{\rm int} = {\rm max}(n,n \pm 1) 
\sigma_{0 \to 1}^{\rm int}$.

At finite temperatures $T$, the deexcitation of the  vibrational number 
$n=1$ to $n=0$ through the inner-band excitation of
the electron-hole pair creation
in the $\pi$-band
can occur. 
In the inner-band excitation, 
both the initial state and the final state of the electron
are  within the states of $| \alpha' \rangle$, which have
dominant component in the $\pi$-band on the surface.
The deexcitation of the vibration with the inner-band excitation
does not occur at 0K.
The rate of the deexcitaion $\sigma_{1 \to 0}^{\rm inn}$ of
vibration through the inner excitation is independent
of $I_{\rm t}$ and depends on $T$:

\begin{eqnarray} 
\sigma_{1 \to 0}^{\rm inn} &=& 2\frac{2\pi}{\hbar} 
\sum_{\alpha'_1, \alpha'_2} |\langle \alpha'_2,0 |H_{\rm e-v} | \alpha'_1,1 \rangle|^2,
\delta(E(\alpha'_2)-\hbar \omega -E(\alpha'_1)) \nonumber \\
&=& \frac{4 |\delta \varepsilon|^2}{\pi \hbar}
\int_{-D_\pi + E_\pi}^{E_\pi-\hbar \omega} {\rm d} \varepsilon \>
\rho_{\rm s}(\varepsilon)\rho_{\rm s}(\varepsilon + \hbar \omega)  \nonumber \\
&& {} \quad \times 
(\frac{1}{\exp(\beta(\varepsilon-\varepsilon_{\rm F}))+1})
(1-\frac{1}{\exp(\beta(\varepsilon + \hbar \omega -\varepsilon_{\rm F}))+1}) \nonumber \\
&\approx& 4\pi |\delta \varepsilon|^2 \rho_{\rm s}^2 (\varepsilon_a)
(\frac{F(T)}{1-\exp(-\beta \hbar \omega)}) \nonumber \\
&\approx& \frac{4}{\pi} W
(\frac{F(T)}{1-\exp(-\beta \hbar \omega)}),   \nonumber \\
F(T) &=& \frac{1}{\beta \hbar} 
\log(\frac{1+\exp(\beta (E_\pi-\varepsilon_{\rm F}))}
{1+\exp(\beta (E_\pi-\hbar \omega -\varepsilon_{\rm F}))}).
\end{eqnarray} 
The transition processes of $\sigma_{\rm elas}$,  
$\sigma_{1 \to 0}^{\rm int}$,  
$\sigma_{0 \to 1}^{\rm int}$,  
$\sigma_{1 \to 0}^{\rm inn}$, and  
$\sigma_{0 \to 1}^{\rm inn}$
are schematically shown in Fig. 1.  
\begin{figure}[hbt]
\begin{center}
\begin{minipage}{0.5\linewidth}
\begin{center}
\includegraphics{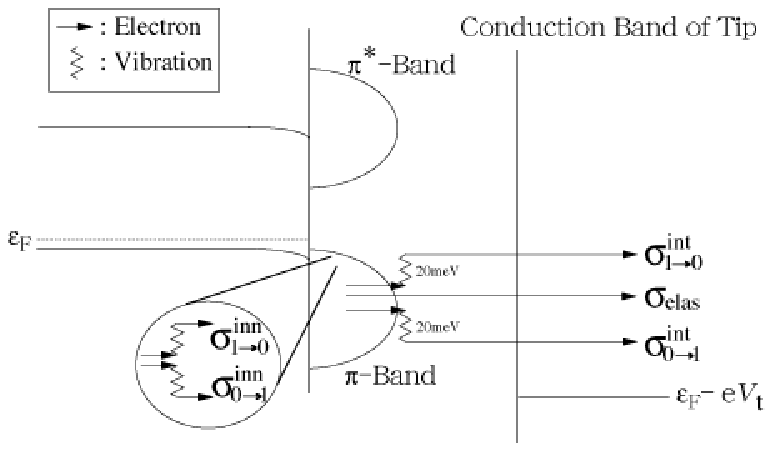}
\end{center}
\caption{
Fig. 1
The transition processes of the inter-band transitions
and the inner-band transitions. 
The rates of the inter-band transitions of 
$\sigma_{\rm elas}$,  
$\sigma_{1 \to 0}^{\rm int}$, and  
$\sigma_{0 \to 1}^{\rm int}$
between the $\pi$-band
and the conduction band of the STM tip
depend on $I_{\rm t}$, and not on $T$.
The inner-band transitions 
of $\sigma_{1 \to 0}^{\rm inn}$ and  
$\sigma_{0 \to 1}^{\rm inn}$
within the $\pi$-band
denpend on $T$, and not on $I_{\rm t}$.
}
\end{minipage}
\end{center}
\end{figure}
The essential framework of the formalism for
$\sigma_{\rm elas}$, 
$\sigma_{0 \to 1}^{\rm int}$, and $\sigma_{1 \to 0}^{\rm int}$
is similar to that of theoretical studies$^{34-41)}$ 
on the STM induced vibration of molecules adsorbed on metal surfaces.
In the problems of molecules adsorbed on metal surfaces,
transitions corresponding to the inner-band transitions in the
present study have also been treated.
The transitions corresponding to the inner-band transitions,
however, do not exhibit any temperature dependence in the case 
of metal surfaces.
Eq. (7) shows that $\sigma_{1 \to 0}^{\rm inn}$ depend strongly 
on $T$; $\pi \beta \hbar \sigma_{1 \to 0}^{\rm inn} \approx 
4W\exp(\beta(E_\pi - \varepsilon_{\rm F}))$ for 
$-\beta(E_\pi - \varepsilon_{\rm F}) \gg 1$
and $\pi \beta \hbar \sigma_{1 \to 0}^{\rm inn} \approx  2W$
for $-\beta(E_\pi -\hbar \omega - \varepsilon_{\rm F}) \ll 1$.
The rate of the excitation $\sigma_{0 \to 1}^{\rm inn}$ of
vibration through the inner-band deexcitation is easily obtained by
the essentially same way as $\sigma_{1 \to 0}^{\rm inn}$:
$\sigma_{0 \to 1}^{\rm inn} \approx (4/\pi) W [F(T)/(\exp(\beta \hbar \omega)-1)]$.

The total transition rate of the excitation $R_{0 \to 1}$ 
of the vibrational number $n=0$ to 1
and
the deexcitation $R_{1 \to 0}$ of the vibrational number $n=1$ to 0
are given by the sum of rates as

\begin{eqnarray} 
R_{0 \to 1} &=& (\frac{\tau_0^{-1}}{\exp(\beta \hbar \omega)-1})
+\sigma_{0 \to 1}^{\rm inn}+\sigma_{0 \to 1}^{\rm int} \nonumber \\
&=& (\frac{1}{\exp(\beta \hbar \omega)-1})
(\tau_0^{-1}+\frac{4}{\pi} W F(T)) + W \sigma_{\rm elas},  \nonumber \\
R_{1 \to 0} &=& (\frac{\tau_0^{-1}}{1-\exp(-\beta \hbar \omega)})
+\sigma_{1 \to 0}^{\rm inn}  +\sigma_{1 \to 0}^{\rm int}  \nonumber \\
&=& (\frac{1}{1-\exp(-\beta \hbar \omega)})
(\tau_0^{-1}+\frac{4}{\pi} W F(T)) + W \sigma_{\rm elas},
\end{eqnarray} 
where $\tau_0$ is the
life time of the vibration through the background phonon system at 0K.

The probability $P_n$ to find the vibrational state at $n$ satisfies
the rate equation represented by $R_{0 \to 1}$ and  $R_{1 \to 0}$:

\begin{eqnarray} 
\frac{{\rm d} P_n}{{\rm d}t} &=& 
(n+1)R_{1 \to 0}P_{n+1}+nR_{0 \to 1}P_{n-1}
-(nR_{1 \to 0}+(n+1)R_{0 \to 1})P_n, \quad 1 \le n,   \nonumber \\
\frac{{\rm d} P_0}{{\rm d}t} &=& R_{1 \to 0}P_1-R_{0 \to 1}P_0.
\end{eqnarray} 
In the stationary distribution of $P_n$, $P_n$ is derived as
$P_n=r^n (1-r)$ from eq. (9), where $r=R_{0 \to 1}/R_{1 \to 0}$.

The tip current $I_{\rm t}$ are given by the sum of the elastic
inter-band transitions and the inelastic inter-band transitions: 

\begin{eqnarray}
\frac{I_{\rm t}}{e}&=&
\sum_{n=0}^\infty P_n \sigma_{\rm elas}
+\sum_{n=0}^\infty P_n \sigma_{n \to n+1}^{\rm int}
+\sum_{n=1}^\infty P_n \sigma_{n \to n-1}^{\rm int} \nonumber \\
&=& [1+W \frac{1+r}{1-r}] \sigma_{\rm elas}.
\end{eqnarray} 

When $T$, $W$, and $I_{\rm t}$ are given, 
$R_{0 \to 1}$ and $R_{1 \to 0}$ are obtained selfconsistently
with eqs. (8) and (10). The distribution of the dimer vibration 
in the STM observation is 
characterized by the effective temperature $T_{\rm ef}$
of the dimer vibration defined as 

\begin{eqnarray}
r=\exp(-\hbar \omega / (k_{\rm B} T_{\rm ef})).
\end{eqnarray} 
Throughout this paper, $T$ denotes the temperature of the substrate.
$T_{\rm ef}$ differs from $T$ in general and depends on
$T$ and $I_{\rm t}$ nonlinearly.
The key quantity for $T_{\rm ef}$ is the fraction 
$W=\sigma_{0 \to 1}^{\rm int}/\sigma_{\rm elas}$,
which is derived
from $|\delta \varepsilon|$ and $\Delta_{\rm s}$ 
as shown in eq. (6).
The coupling constant $\delta \varepsilon$ 
between the amplitude of the vibration and the energy of $| a \rangle$,  
is approximately estimated as

\begin{eqnarray}
|\delta \varepsilon| &=& \Delta \theta \frac{\partial \varepsilon_a}{\partial \theta}_{\theta=\theta_0},
\end{eqnarray} 
where $\theta_0$ (= $19.1^\circ = 0.333$rad ) is the equilibrium tilting angle of dimer 
in the c(4 $\times$ 2) structure, and $\Delta \theta$ is 
the mean value of 
the amplitude of the vibrational number $n=1$.
$\Delta \theta$ is obtained by $\omega$ and the effective value of
the moment of inertia $I$ as $\Delta \theta = (3\hbar/(2 I \omega))^{1/2}$.
As mentioned already, the model potentials for the continuous
values of the tilting angles of the dimers on Si(001) surface
are obtained based on the FPC results. 
We can get approximate values 
of $I$ and $\partial \varepsilon_a / \partial \theta$ 
at $\theta=\theta_0$ by the model potentials: 

\begin{eqnarray}
\frac{\partial \varepsilon_a}{\partial \theta}_{\theta=\theta_0} &=& - k \theta_0, 
\nonumber \\
I &=& \frac{k}{\omega^2}, 
\end{eqnarray} 
where $k$ is
the value of the second partial derivative of the model potentials 
for the tilting angle of the dimer beneath the tip at $\theta=\theta_0$.
We derive the value of $k$ from the model potentials, as
$k=1.1 \times 10^4{\rm meV}/{\rm rad}^2$.
Using eq. (13), we get approximate values as 
$I=1.2 \times 10^{-23}{\rm meVs}^2/{\rm rad}^2$,
$\partial \varepsilon_a / \partial \theta =
 -3.7 \times 10^3{\rm meV}/{\rm rad}$, 
and $\Delta \theta = 5.2 \times 10^{-2}{\rm rad}$.
From eq. (12), $|\delta \varepsilon|$ is approximately estimated 
to be $1.9 \times 10^2{\rm meV}$.
From the half width of half maximum of STS for $\pi$-band of Si(001)
at 5.5K, $\Delta_{\rm s}$ is roughly estimated to be $2.5 \times 10^2$meV.
From these values, 
the key quantity $W$ 
is obtained to be about 0.6.\\

{\parindent=0pt \bf 3. Results and Discussions}

In the present study, we take the scope on the p-type substrate.
The Fermi level of the  p-type substrate is well approximated to be 
at the center
of the top of the valence band of the substrate and the acceptor level
of dopant at low temperatures.
At temperatures low but higher than a few Kelvin,
$E_\pi$ is practically fixed at the top of the valence band.
In the present study, $E_\pi - \varepsilon_{\rm F}$ is approximated
to be $-22.5$meV for B-doped substrate and  
$-32.5$meV for Ga-doped substrate, respectively.
The life time of the vibration $\tau_0$ through the background
phonon system at 0K is expected to be of
the order of $10^{-9}$s for semiconductors.
We assumed for $\tau_0$ to be 8ns.

The excitation rate $R_{0 \to 1}$ and 
the deexcitation rate $R_{1 \to 0}$ for B-doped substrate
are shown in Fig. 2.
\begin{figure}[hbt]
\begin{center}
\begin{minipage}{0.97\linewidth}
\begin{center}
\includegraphics[width=16cm]{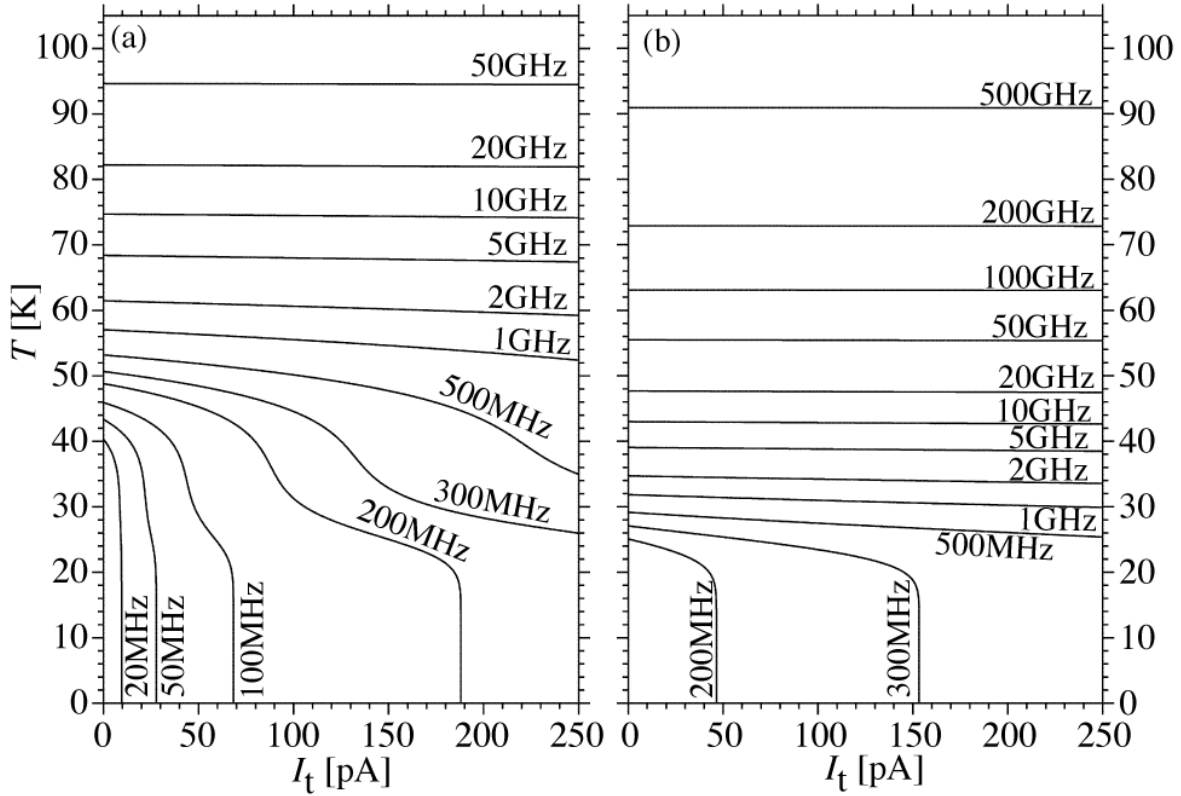}
\end{center}
\caption{
Fig. 2 The total rates for the excitation of the dimer vibration
and  the deexcitation for B-doped substrate.
The rates are obtained selfconsistently with eqs. (8)
and (9). 
(a): the excitation rate $R_{0 \to 1}$. 
(b):  the dexcitation rate $R_{1 \to 0}$. 
}
\end{minipage}
\end{center}
\end{figure}
Both rates are monotonically increasing function of 
$T$ and $I_{\rm t}$ in highly nonlinear manner.  
The rates scarcely depend on $T$ for $T<20$K, 
because the inter-band transitions are dominant
in them and the inner-band transitions scarcely contribute 
to them. 
When $T$ increases around 20K,
the excitation and deexcitaion of the vibration
through the inner-band transitions of electrons
become effective and the rates increase steeply for $T$.
The deexcitation rate $R_{1 \to 0}$
increases much steeper for $T$ than $R_{0 \to 1}$
around 20K $\sim$ 40K.
The rates scarcely depend on $I_{\rm t}$ for $T>70$K, 
because the inner-band transitions are dominant
in them and the inter-band transitions scarcely contribute 
to them. 

Since the magnitudes of the rates are smaller enough than
the frequency of the vibration 4.8THz derived from $\hbar \omega =
20$meV, 
the effective temperature of the dimer vibration
$T_{\rm ef}=-\hbar \omega / (k_{\rm B} \log(R_{0 \to 1}/R_{1 \to 0}))$ 
well characterizes the distribution of the vibration.
$T_{\rm ef}$ for B-doped substrate are shown in Fig. 3.
\begin{figure}[hbt]
\begin{center}
\begin{minipage}{0.7\linewidth}
\begin{center}
\includegraphics[width=8.5cm]{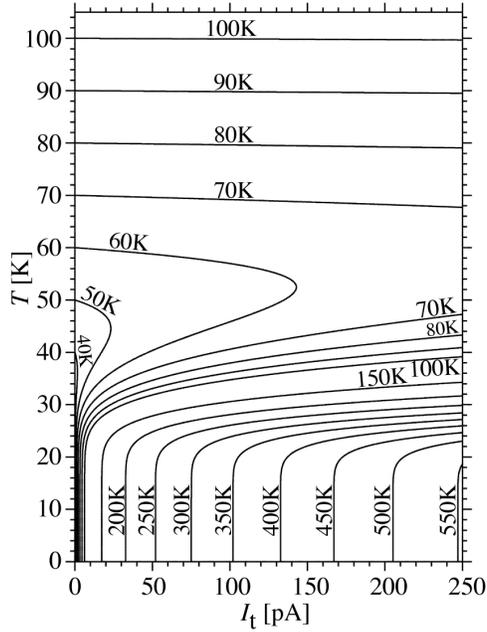}
\end{center}
\caption{
Fig. 3 The effective temperature of the dimer vibration ($T_{\rm ef}$)
for B-doped substrate. $T_{\rm ef}$ reaches 
about 500K at $I_{\rm t}=200$pA at low temperature $T < 20$K.
$T_{\rm ef}$ decreases steeply around 
$T \approx 20K \sim 40K$, as $T$ increases.  
}
\end{minipage}
\end{center}
\end{figure}
At low temperatures of $T<20$K, 
$T_{\rm ef}$ scarcely depending on $T$, increases steeply 
with $I_{\rm t}$, becomes about 250K at $I_{\rm t}=50$pA,
and reaches about 500K at $I_{\rm t}=200$pA.
The rapid flip-flop is expected in such high $T_{\rm ef}$.
These high $T_{\rm ef}$ appearing at low temperatures of $T<20$K,
essentially explains why the symmetric STM images are
observed at low temperatures.
When $T$ increases around 20K, $T_{\rm ef}$ starts to decrease
steeply, because of steep increase of $R_{1 \to 0}$.
$T_{\rm ef}$ becomes 
70K around $T \approx 40$K and $I_{\rm t} \approx 150$pA.
At higher temperatures $T>60$K, 
$T_{\rm ef}$ increases gradually with $T$, 
scarcely depends on $I_{\rm t}$ 
within the typical values of $I_{\rm t}$,
and takes the nearly same value as $T$. 
These low $T_{\rm ef}$  explains why the asymmetric STM
images recover in temperature range higher than 50K.

$T_{\rm ef}$ for Ga-doped substrate are shown in Fig. 4.
The essential feature of  $T_{\rm ef}$ shown in Fig. 4
is same as in Fig. 3. The steep decrease of $T_{\rm ef}$
in Fig. 4,  
however, is starting around $T \approx 30$K
which is  higher
than for B-doped substrate shown in Fig. 3.
The temperature where the inner-band transitions
exceed the inter-band transitions in magnitude on Ga-doped substrate, 
is higher than that on B-doped substrate 
and the ratio of these two temperatures is almost the same
value as the ratio of $\varepsilon_{\rm F}-E_\pi$,
32.5/22.5=1.44,
since $\sigma_{1 \to 0}^{\rm inn}$, as mentioned already,
is approximated as 
$\sigma_{1 \to 0}^{\rm inn} \approx 
4W \exp(\beta(E_\pi-\varepsilon_{\rm F}))/(\pi \beta \hbar)$ 
for $-\beta(E_\pi-\varepsilon_{\rm F}) \gg 1$.
\begin{figure}[hbt]
\begin{center}
\begin{minipage}{0.7\linewidth}
\begin{center}
\includegraphics[width=8.5cm]{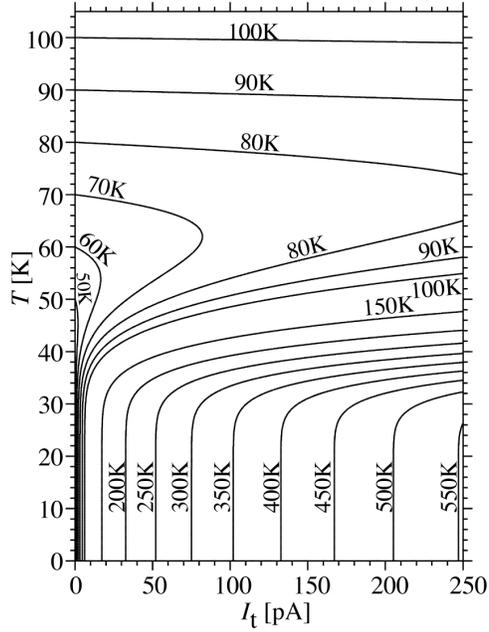}
\end{center}
\caption{
Fig. 4 The effective temperature of the dimer vibration ($T_{\rm ef}$)
for Ga-doped substrate. $T_{\rm ef}$ reaches 
about 500K at $I_{\rm t}=200$pA at low temperature $T < 30$K.
$T_{\rm ef}$ decreases steeply around 
$T \approx 30K \sim 60K$, as $T$ increases.  
The decreases on Ga-doped substrate appears in higher temperatures
than on B-doped substrate.
}
\end{minipage}
\end{center}
\end{figure}

At low temperature $T<20$K,
the atom at the up-position beneath the tip is rapidly
transferred to the
down-position through the dimer flip-flop motion
by quasi-thermal excitation of the vibration of high $T_{\rm ef}$.
After the flip-flop motion to the down-position,
$T_{\rm ef}$ becomes much lower than at the up-position,
because tip current is expected to be much smaller at down-position
than at up-position. 
Because
of the low effective temperature $T_{\rm ef}$,
the transition rate of the atom from the down-position
to the up-position is much smaller than that of the opposite transition;
the fraction of the staying time at the down-position is much
larger than that at the up-position.
The spatial STM gap between
the prominence of the tip and the surface dimer
is tuned for $I_{\rm t}$ to be the STM condition.
The tip current $I_{\rm t}$ is time averaged one, and
is tuned almost at the down-positon at low temperatures.
In the scanning of STM tip at much low temperatures, 
the atoms beneath the tip are almost always
tied to the down-positions, and the atoms at the up-position
scarcely contribute to the STM images.
As the result of the large fraction at the down-position, 
the symmetric dimer images  
of the p(2 $\times$ 1) structure is observed
even in the ordered state of c(4 $\times$ 2).
These feature that the tip ties the atom to the down-positions
in STM observation on Si(001) surface
corresponds to the experimental results
reported by Mitsui and Takayanagi$^{42)}$.

In the present study, we demonstrate that the inner-band transitions
play the important role to decrease $T_{\rm ef}$ 
at the temperatures
of a few ten Kelvin
where  deexcitation of the vibration through 
the inner-band excitations of the electron-hole creation
occur.
This deexcitation through the inner-band excitation
can occur also when enough amount of holes in the $\pi$-band
or enough amount of electrons in the $\pi^*$-band is induced.
For example,
when the electrons and holes are induced by infrared photons or by other
methods,
$T_{\rm ef}$ will be expected to decrease 
in much low temperature in STM observation.
Here we propose the notion of ``the excitation-induced cooling'', namely,
the decrease of $T_{\rm ef}$  essentially
induced by the electronic excitation in $\pi$-band or $\pi^*$-band.\\

{\parindent=0pt \bf 4. Conclusion}

In order to explain the symmetric-asymmetric crossover in dimer
image of STM observation on Si(001) surface,
we calculate the effective temperature $T_{\rm ef}$ of 
dimer vibration.
The crossover behavior is semiquantitatively
understood by the nonlinear dependence of $T_{\rm ef}$
on the substrate temperature and STM current.
The symmetric image 
in the ordered state of c(4 $\times$ 2)
results from 
the large fraction of the atom at the down-position
induced by STM current.
We have proposed the notion of ``the excitation-induced cooling''
under more general situations than 
that induced by the substrate temperature.\\


{\parindent=0pt \bf  Acknowledgment}

The author would like to thank K. Makoshi, H. Ueba, and T. Mii
for valuable discussions.
This work was supported in part 
by Grant-in-Aid for Scientific Research (C)
from Japan Society for the Promotion of Science.

\newpage

{\parindent=0pt \bf References}\\
{\parindent=0pt
\begin{tabular}{rl}
$1)$& R. E. Schlier and H. E. Farnsworth: J. Chem. Phys. {\rm 30} (1959) 917. \\
$2)$& J. J. Lander and J. Morrison: J. Chem. Phys. {\rm 37} (1959) 729. \\
$3)$& T. Tabata, T. Aruga, and Y. Murata: Surface Science {\bf 179} (1987) L63. \\
$4)$& R. J. Hamers, R.M. Tromp, and J. E. Demuth: 
           Phys. Rev. B {\bf 34} (1986) 5343. \\
$5)$& R. A. Wolkow: Phys. Rev. Lett. {\bf 68} (1992) 2636. \\
$6)$& K. Inoue, Y. Morikawa, K. Terakura, and M. Nakayama: \\
    &  Phys. Rev. B {\bf 49} (1994) 14774. \\
$7)$& K. Terakura, T. Yamasaki, and Y. Morikawa: 
           Phase Transition {\bf 53} (1995) 143. \\
$8)$& S. D. Kevan and N. G. Stoffel: Phys. Rev. Lett. {\bf 53} (1984) 702. \\
$9)$& S. D. Kevan: Phys. Rev. B {\bf 32} (1985) 2344. \\
$10)$& S. Ferrer, X. Torrelles, V. H. Etgens, H. A. van der Vegt, and P. Fajardo: \\ 
     & Phys. Rev. Lett. {\bf 75} (1995) 1771.\\
$11)$& C. A. Lucas, C. S. Dower, D. F. McMorrow, G. C. L. Wong, F. J. Lamelas, and \\
     & P. H. Fuoss: Physi. Rev. B {\bf 47} (1993) 10375.\\
$12)$& T. Sato, M. Iwatsuki, and H. Tochihara: 
      J. Electron Microsc. {\bf 48} (1999) 1. \\
$13)$& Y. Yoshimoto, Y. Nakamura, H. Kawai, M. Tsukada, and M. Nakayama: \\ 
     & Phys. Rev. B {\bf 61} (2000) 1965.\\
$14)$& Y. Yoshimoto, Y. Nakamura, H. Kawai, M. Tsukada, and M. Nakayama: \\
     & Surf.  Rev. and Lett. {\bf 6} (1999) 1045.\\
$15)$& H. Shigekawa, K. Miyake, M. Ishida, K. Hata, H. Oigawa, Y. Nannichi, \\ 
     & R. Yoshizaki, A. Kawazu, T. Abe, T. Ozawa, and T. Nagamura: \\
     & Jpn. J. Appl. Phys. {\bf 35} (1996) L1081.\\
$16)$& S. Yoshida, O. Takeuchi, K. Hata, R. Morita, M. Yamashita, 
      and H. Shigekawa: \\
     & Jpn. J. Appl. Phys. {\bf 41} (2002) 5017.\\
$17)$& K. Hata, S. Yoshida, and H. Shigekawa: Phys. Rev. Lett. {\bf 89} (2002) 286104.\\  
$18)$& M. Matsumoto, K. Fukutani, and T. Okano: Phys. Rev. Lett. {\bf 90} (2003) 106103.\\  
$19)$& T. Yokoyama and K. Takayanagi: Phys. Rev. B {\bf 61} (2000)
 R5078.\\
$20)$& K. Kondo, T. Amakusa, M. Iwatsuki, and H. Tokumoto: Surface
 Science {\bf 453} (2001) L318. \\ 
$21)$& A. Ramstad, G. Brocks, and P. J. Kelly: Phys. Rev. B {\bf 51}
 (1994) 14504. \\
$22)$&  H. Tochihara, Y. Nakamura, H. Kawai, M. Nakayama, T. Sato, T. Sueyoshi, \\
       & T. Amakusa, and M. Iwatsuki: J. Phys. Soc. Jpn. {\bf 67} (1998) 2330. \\
$23)$& Y. Nakamura, H. Kawai, and M. Nakayama:
     Phys. Rev. B {\bf 52}52 (1995) 8231.\\
\end{tabular}
\begin{tabular}{rl}
$24)$& Y. Nakamura, H. Kawai, and M. Nakayama:
     Surface Science {\bf 357-358} (1996) 500. \\
$25)$& Y. Nakamura, H. Kawai, and M. Nakayama:
       Phys. Rev. B {\bf 55} (1997) 10549.\\
$26)$& H. Kawai, Y. Nakamura, and M. Nakayama:
   J. Phys. Soc. Jpn. {\bf 68} (1999) 3936. \\
$27)$& H. Kawai, R. Miyata, Y. Yoshimoto, and M. Tsukada: submitted to 
J. Phy. Soci. Japn. \\
$28)$& H. Kawai, Y. Yoshimoto, H. Shima, Y. Nakamura, and Masaru Tsukada:\\
     &  J. Phys. Soc. Jpn {\bf 71} (2002) 2192.\\
$29)$& K. Hata, Y. Saino, and H. Shigekawa: Phys. Rev. Lett. {\bf 86}
 (2001) 3084.\\ 
$30)$& Y. Enta, S. Suzuki, and S. Kono: Phys. Rev. Lett. {\bf 65}
 (1990) 2704.\\
$31)$& T.Yokoyama, M. Okamoto, and K. Takayanagi: Phys. Rev. Lett. {\bf 81}
 (1998) 3423.\\
$32)$& K. Hata, Y. Shibata, and H. Shigekawa: Phys. Rev. B {\bf 64}
 (2001) 235310.\\   
$33)$ &  N. Takagi, S. Shimonaka, T. Aruga, and N. Nishijima: 
         Phys. Rev. B {\bf 60} (1999) \\
      & 10919. \\
$34)$& B. N. J. Persson and M. Persson: Solid State Comm. {\bf 36}
(1980) 175. \\
$35)$& B. N. J. Persson and A. Baratoff: Phys. Rev B {\bf 59} (1987) 339.\\ 
$36)$& S. Gao, M. Persson, and B. I. Lundqvist: Phys. Rev. B {\bf 55} (1997)
4825. \\
$37)$& N. Mingo and K. Makoshi: Surface Science {\bf 438} (1999) 261. \\
$38)$& N. Mingo and K. Makoshi: Phys. Rev. Lett. {\bf 84} (2000) 3694. \\
$39)$& K. Makoshi, N. Mingo, T. Mii, H. Ueba, and S. Tikhodeev:
Surface Science {\bf 493} (2001) 71.\\
$40)$& N. Mingo, K. Makoshi, T. Mii, and H. Ueba: Surface Science {\bf 482-485}
(2001) 96. \\
$41)$& B. N. J. Persson, H. Ueba: Surface Science {\bf 502-503} (2002) 18.\\
$42)$& T. Mitsui and K. Takayanagi: Phys. Rev. B {\bf 62} (2000) R16251. \\
\end{tabular}
}

\end{document}